\documentclass[prb,aps,amssymb,twocolumn,showpacs,floatfix]{revtex4}
\usepackage{graphics}
\usepackage{epsfig} 
\usepackage[dvips]{color}
\usepackage{dcolumn}
\usepackage{amsmath}
\usepackage{color}
\begin{document}
\newcommand\dd{{\operatorname{d}}}
\newcommand\sgn{{\operatorname{sgn}}}
\def\Eq#1{{Eq.~(\ref{#1})}}
\def\Ref#1{(\ref{#1})}
\newcommand\e{{\mathrm e}}
\newcommand\cum[1]{  {\Bigl< \!\! \Bigl< {#1} \Bigr>\!\!\Bigr>}}
\newcommand\vf{v_{_\text{F}}}
\newcommand\pf{p_{_\text{F}}}
\newcommand\ef{{\varepsilon} _{\text{\sc f}}}
\newcommand\zf{z_{_\text{F}}}
\newcommand\zfi[1]{{z_{_\text{F}}}_{#1}}
\newcommand\av[1]{\left<{#1}\right>}
\def\det{{\mathrm{det}}}
\def\Tr{{\mathrm{Tr}}}
\def\Li{{\mathrm{Li}}}
\def\tr{{\mathrm{tr}}}
\def\im{{\mathrm{Im}}}
\def\Texp{{\mathrm{Texp}\!\!\!\int}}
\def\antiTexp{{\mathrm{\tilde{T}exp}}\!\!\!\int}
\title{Non-equilibrium 1D many-body problems and asymptotic properties of
Toeplitz determinants} 

\author{D. B. Gutman$^{1}$, Yuval Gefen$^2$, and A. D. Mirlin$^{3,4,5}$}
\affiliation{
\mbox{$^1$Department of Physics, Bar Ilan University, Ramat Gan 52900,
Israel }\\
\mbox{$^2$Dept. of Condensed Matter Physics, Weizmann Institute of
  Science, Rehovot 76100, Israel}\\
\mbox{$^3$Institut f\"ur Nanotechnologie, Karlsruhe Institute of Technology, 
 76021 Karlsruhe, Germany}\\
\mbox{$^4$Institut f\"ur Theorie der kondensierten Materie,
Karlsruhe Institute of Technology, 76128 Karlsruhe, Germany}\\
\mbox{$^5$Petersburg Nuclear Physics Institute, 188300 St.~Petersburg, Russia}
}

\date{\today}

\begin{abstract}
Non-equilibrium bosonization technique facilitates the  solution of a number
of  important  many-body problems out of equilibrium, including the Fermi-edge
singularity, the tunneling spectroscopy and full counting statistics of
interacting fermions forming a Luttinger liquid. 
We generalize the method to non-equilibrium
hard-core bosons (Tonks-Girardeau gas) and establish interrelations
between all these problems. The results 
can be expressed in terms of  Fredholm determinants of Toeplitz type.
We analyze  the  long time asymptotics of such determinants, using
Szeg\H{o} and Fisher-Hartwig theorems. Our analysis 
yields dephasing rates as well
as power-law scaling behavior, with exponents depending not only on the
interaction strength but also on the non-equilibrium state of the
system.
\end{abstract}
\maketitle

\section{Introduction}
There is a number of quantum many-body problems that are of paramount
importance for condensed matter physics and, at the same time,
possess an exact solution. 
These are the Anderson orthogonality catastrophe\cite{Anderson}, Fermi edge
singularity\cite{Nozieres} (FES), Luttinger liquid\cite{Tomonago} (LL)
zero-bias anomaly\cite{Luther}, and Kondo problems\cite{Kondo}.  
It has been realized long ago   that  these problems  are,  in fact,
inter-connected,   
both vis-a-vis   the  underlying physics and the  mathematics  involved.
Such connections have been used, e.g.,  for the
representation of  the dynamics of the  Kondo problem  
as  an infinite sequence of  Fermi-edge-singularity events\cite{Yuval}. 

In our recent works\cite{GGM_long2010,GGM_short2010}  
we have addressed some of these problems away 
from equilibrium.  To achieve this goal, 
we  have developed a non-equilibrium bosonization technique
generalizing the conventional bosonization
\cite{stone,Delft,Gogolin,giamarchi,maslov-lectures} onto problems
with non-equilibrium distribution functions.
We have shown that the relevant correlation functions  can be
expressed through  
Fredholm determinants of a ``counting'' operator. 
The information on the specific type of the problem, as well as on different 
aspects of the interaction, 
is  encoded  in the time-dependent scattering phase of the counting operator.  
The results of 
Refs.~\onlinecite{GGM_long2010,GGM_short2010} have demonstrated  that those 
classic many-body problems are more closely connected than has been 
previously understood, extending the interrelations into  the
non-equilibrium regime.  

The purpose of the present  paper is twofold.
First, we present the  relation between the LL tunneling
spectroscopy, the full 
counting statistics, and the  Fermi-edge singularity problems in a
systematic way. We also include in this scheme the generalization of
the LL problem to a non-equilibrium gas of  strongly repulsive
bosons (Tonks-Girardeau gas). These results are summarized in
Table~\ref{table1}.  
Second, we apply the general theory of
asymptotic behavior of 
Toeplitz determinants to the present problems. This allows us to
extract important  information on the  many-body physics of these models,
including non-equilibrium dephasing rates and 
the modification of power-law exponents due to
non-equilibrium conditions. 

\setlength{\tabcolsep}{7pt}
\begin{table}
\begin{center}
\resizebox{9cm}{!}{
  \begin{tabular}{ | c  | c |  c |  c | c | }
    \hline 
       & $c_R$ & $c_L$ & $\delta_R$ & $\delta_L$ \\ \hline \hline
    $G_{\rm FES}^>$ & $-1+
\frac{\delta_0}{\pi}$  & $0$ &  $2(\pi -\delta_0)$ & $0$ \\ \hline
    $G_{FR}^>$  &  $-1$  &  0  &  $2\pi \frac{1+K}{2\sqrt{K}}$   &
    $2\pi\frac{1-K}{2\sqrt{K}}$  \\ \hline  
   $G_{FL}^>$  & 0 & $-1$ & $2\pi\frac{1-K}{2\sqrt{K}}$ &
   $2\pi\frac{1+K}{2\sqrt{K}}$ \\ \hline 
   $\chi_\tau(\lambda)$ & $-\frac{\lambda}{2\pi}$ &
   $\frac{\lambda}{2\pi}$ & $\lambda\sqrt{K}$ & $-\lambda\sqrt{K}$ \\ 
    \hline
$G_B^>$& $-\frac{1}{2}$ & $-\frac{1}{2}$ & $\frac{\pi}{\sqrt{K}}$ &
$\frac{\pi}{\sqrt{K}}$ \\ 
\hline
 \end{tabular}
} 
\label{table1} 
\caption{Non-equilibrium correlation functions of many-body
  problems: Fermi edge singularity ($G^>_{\rm FES}$),  Green functions
  of right- and left-moving fermions in a LL ($G^>_{\rm FR}$ and
  $G^>_{\rm FL}$), full counting statistics of a LL
  ($\chi_\tau(\lambda)$), Green function of the Tonks-Girardeau gas
  ($G^>_{\rm B}$). All these correlation functions can be cast in the
  form $G^> = \langle e^{-i{\cal O}_-(\tau)}e^{i{\cal O}_+(0)}\rangle$,
  where ${\cal O} = \sum _{\eta=R,L} c_\eta\phi_\eta$ and $\phi_\eta$
  are free bosonic fields. The coefficients $c_\eta$ are shown in the
  second and third columns of the table.
Evaluation of these correlation functions
  yields the results in the form of Fredholm-Toeplitz determinants
  $\Delta_R[\delta_R]\Delta_L[\delta_L]$. The corresponding phases
  $\delta_{R,L}$ are presented in the last two columns for the case
  where the LL is adiabatically connected to the  reservoirs.} 
\end{center}
\end{table}

The structure of the present  paper is the  following:
In Sec.~\ref{fermionic_problems} we  summarize the results  of our
earlier works \cite{GGM_long2010,GGM_short2010} for the FES problem,
as well as the
tunneling spectroscopy and the  counting statistics problems  in the LL, all 
underlined  
by interacting fermions.  
Section \ref{bosonic_problems} is devoted to the
tunneling spectroscopy problem  for a non-equilibrium Tonks-Girardeau gas. 
In Sec.~\ref{Fredholm_determinant} we  study the long-time behavior of
emerging Fredholom determinants, employing   Szeg\H{o} and
Fisher-Hartwig theorems.  For completeness, properties of the  
Tonks-Girardeau gas at
equilibrium are 
summarized  in Appendix \ref{Tonks_girardo_equilibrium}, while 
relevant mathematical  theorems from the theory of Fredholm determinants
 are  presented  
 in Appendix \ref{Fisher-Hartwig-conjecture}.

\section{Many-body problems as Fredholm Determinants}

\subsection{Fermionic problems}
\label{fermionic_problems}

\subsubsection{Free fermions}
\label{free_fermions}

We first discuss   non-interacting electrons.
Evidently,  the  Green functions of free fermions $G_0^>$, $G_0^<$ (we
use the standard notations of Keldysh formalism \cite{Kamenev} 
can easily  be  calculated within a fermionic description.  
However,  it is insightful to recalculate it  employing the
non-equilibrium bosonization\cite{GGM_long2010}. The result reads 
\begin{equation}
\label{a1}
G_{0,\eta}^\gtrless(\tau)=-\frac{1}{2\pi v}\frac{1}{\tau\mp i/\Lambda}
\overline{\Delta}_\eta[\delta_\eta(t)]\:.
\end{equation}
Here  $\eta$ is the chirality index ($\eta=L$ for left and $\eta=R$
for right movers), the phase $\delta_\eta(t)=\lambda\omega_\tau(t,0)$,
where
\begin{equation}
w_\tau(t,\tilde{t})=\theta(\tilde{t}-t)-\theta(\tilde{t}-t-\tau)
\end{equation}
is a ``window function'', and
$\lambda=2\pi$. Further,
$\Delta_\eta[\delta_\eta(t)]$  is 
the Fredholm determinant of a counting operator
\begin{equation}
\label{determinant_definition}
\Delta_\eta[\delta_\eta(t)]=
\det[1
  + (e^{-i{\hat{\delta}_\eta}}-1)\hat{n}_\eta]\,, 
\end{equation}
and we denote by $\overline{\Delta}_\eta$ the determinant normalized
to its value for zero-$T$ equilibrium distribution. 
As we see, in the bosonic description, a  free fermion is represented
by a $2\pi$ phase soliton entering the counting operator.
The latter consists of a  fermionic distribution function $n_\eta(\epsilon)$,  
and  
a time-dependent scattering phase $\delta_\eta(t)$, with $\epsilon$
and $t$ to be understood as canonically conjugate variables.
Since the matrices $\hat{\delta}$ and $\hat{n}$ 
can not be diagonalized simultaneously,
the task of calculating   $\Delta_\eta[\delta_\eta(t)]$ is non-trivial.
In most cases the analysis of such determinants can be done only numerically.
However, the long time asymptotic properties can be studied analytically, 
by means of  Szeg\H{o} and Fisher-Hartwig theorems, 
as discussed in Sec.~\ref{Fredholm_determinant}.

When read from right to left, Eq.~(\ref{a1}) can be viewed as a
remarkable identity for Fredholm determinants. It yields an exact
value of such a determinant for an  arbitrary distribution function
$n_\eta(\epsilon)$, and for the phase function $\delta(\tau)$ given by
a window function with amplitude $2\pi$. It is worth mentioning that
the determinat should be considered as analytically continued from the
region of small phases $\delta$.

\subsubsection{Fermi edge singularity}

The FES problem describes the  scattering of conduction  electrons off a
localized hole, which  is left behind by  an electron excited into the
conduction band. 
Historically, the FES problem was first solved by exact summation of an
infinite diagrammatic series \cite{Nozieres}.
While in the FES problem there is no interaction between electrons
in the conducting band, it has many features characteristic of genuine
many-body physics. 
Despite the fact that conventional experimental realizations of FES are
three-dimensional, the problem can be reduced (due to the local and isotropic 
character of the interaction with the core hole) to that of
one-dimensional chiral fermions. For this reason,  bosonization
technique can be effectively applied, leading
to  an alternative and very elegant solution \cite{Schotte}.

One can consider the FES out of equilibrium\cite{Abanin}, 
with an arbitrary electron distribution function $n(\epsilon)$. 
This problem can be solved within the framework 
of non-equilibrium bosonization\cite{GGM_long2010}, with the following
results for the  emission/absorption rates: 
\begin{equation}
\label{b1}
G^\gtrless_{\rm FES}(\tau)=\mp
{ i\Lambda{\overline{\Delta}}_{\tau}(2\pi-2\delta_0) \over
2\pi v (1\pm i\Lambda\tau)^{(1-\delta_0/\pi)^2}}.
\end{equation}
Here $\delta_0$ is the $s$-wave electronic phase shift  due to the
scattering of conduction electrons off the core hole.

\subsubsection{Luttinger liquid: Tunneling spectroscopy}

The tunneling spectroscopy technique allows one to explore
experimentally Keldysh
Green functions of an interacting system that carry information about
both tunneling density of states and energy distribution. 
Recent experiments on carbon nanotubes and quantum Hall edges have
proved the efficiency of this technique in the context of 1D
systems\cite{tunnel-spectroscopy,tunnel-spectroscopy-qhe}. The
technological and experimental 
advances motivate the theoretical interest in the tunneling
spectroscopy of strongly correlated 1D structures away from
equilibrium
\cite{GGM_long2010,jakobs08,gutman08,trushin08,pugnetti09,Ngo,takei10,bena10}. 

In the case
of a  LL made of 1D interacting fermions, the Keldysh Green function may be
evaluated theoretically via the non-equilibrium bosonization technique. 
Assuming that a long LL conductor is coupled to
two reservoirs (modeled as non-interacting 1D
wires\cite{Maslov,Safi,Ponomarenko}) with 
distribution functions $n_R(\epsilon)$ and  $n_L(\epsilon)$ respectively, 
one obtains 
\cite{GGM_long2010}   
\begin{equation}
\label{d3}
G^\gtrless_R(\tau)=\mp \frac{i\Lambda}{2\pi u}
\frac{\overline{\Delta}_R[\delta_R(t)]\overline{\Delta}_L[\delta_L(t)]}
{(1\pm i\Lambda \tau)^{1+\gamma}},
\end{equation}
where $u=v/K$ is the sound velocity,
\begin{equation}
\gamma=(1-K)^2/2K\,,
\end{equation}
and
\begin{equation}
K=(1+g/\pi v)^{-1/2}
\end{equation}
 is the standard LL parameter in the interacting region.
 The phase  $\delta_\eta(t)$ is found to be a
 superposition of rectangular pulses,
\begin{equation}
\label{delta_spinless}
\delta_\eta(t)=\sum_{n=0}^\infty \delta_{\eta,n}w_\tau(t,t_n)\,,
\end{equation}
where
\begin{equation}
t_n=(n+1/2-1/2K)L/u\,
\end{equation}
and
\begin{eqnarray}&&
\label{delta_spinless1}
\delta_{\eta,2m}=\pi t_{-\eta} r_L^mr_R^m(1+\eta K)/\sqrt{K}\,,
\nonumber \\&&
\delta_{\eta,2m+1}=
-\pi t_{-\eta} r_\eta^{m+1} r_{-\eta}^{m} (1-\eta K)/\sqrt{K}\,.
\end{eqnarray}
Here $r_\eta$, $t_\eta$ are reflection and transmission coefficients
of plasmons at the left ($\eta=L\equiv -1$) and right
($\eta=R\equiv +1$) boundaries. 
For $\tau \ll L/u$ the coherence of plasmon scattering  may  be
neglected and the result assumes the form of  a product 
\begin{equation}
\label{c10}
\overline{\Delta}_\eta[\delta_\eta(t)]\simeq\prod_{n=0}^\infty
\overline{\Delta}_{\eta\tau}(\delta_{\eta,n})
\,.
\end{equation}
The  plasmon scattering causes a fractionalization ({\it cf.}
Refs.~\onlinecite{Safi,lehur,Safi97,Pham00,LeHur08,Berg09,Deshpande10}) 
of the phase soliton,
splitting it into an infinite series of pulses.
As a result, the Fredholm determinant of a counting operator
takes the form of  an infinite product  of determinants, 
each calculated for  a rectangular pulse with a corresponding scattering phase
$\delta_{\eta,n}(t)=\delta_{\eta,n}w_\tau(t,0)$. 

In the case of smooth boundaries with the  leads we have $r_\eta \simeq
0$, which implies  that only the first ($n=0$) pulse survives in both $\delta_R$
and $\delta_L$. The corresponding amplitudes are depicted in
Table~\ref{table1}. 

The above results correspond to the case where  the tunneling
spectroscopy point is located inside the interacting part of the
wire. The case of tunneling into one of non-interacting leads is
studied in the same way; the results can be found in
Ref.~\onlinecite{GGM_long2010}. 

So far we have discussed Green functions at coinciding spatial points
having in mind tunneling spectroscopy experiments. Green functions at
different spatial coordinates are also of interest, in particular, in
the context of Aharonov-Bohm interferometry\cite{lehur,GGM_long2010}.
(A similar problem in the context of chiral edge states has been
considered in Refs.~\onlinecite{Chalker07,Levkivskyi09,Kovrizhin10}.)
Equation (\ref{d3}) expressing the non-equilibrium LL Green function in
terms of Fredholm determinants can be generalized for this case as
well \cite{GGM_long2010}.

\subsubsection{Luttinger liquid: Counting statistics}

In  the problem of counting statistics of non-equilibrium LL one is
interested in the   generating function  
$\chi(\lambda)=\sum_{n=-\infty}^\infty p_\tau(n) e^{in\lambda}$, where  
$p_\tau(n)$  is  the  probability for $n$ electrons to pass through  a
given  cross-section  during  
the time interval $\tau$. On the experimental side, the second moment
(shot noise) has been measured in correlated 1D systems
\cite{Kim07,Wu07}; an experimental analysis of higher moments
and of the full counting statistics (FCS) remains a challenging issue.   

For the non-interacting case the FCS 
generating function $\chi(\lambda)$ 
has been calculated in Ref.~\onlinecite{Levitov-noise}  
by means of Landauer scattering-state approach. 
For an ideal quantum wire  (with no scattering inside the wire) with
distributions $n_\eta(\epsilon)$,  the
generating function of FCS  is given by  
\begin{equation}
\label{generating_function}
\chi(\lambda)=\Delta_R[\delta_R(t)]\Delta_L[\delta_L(t)]\,,
\end{equation}
with the phases $\delta_\eta(t) = \lambda\eta w_\tau(t,0)$.  
For an interacting wire (LL) the result 
\cite{GGM_short2010} retains the form of a product of Fredholm determinants,
Eq.~(\ref{generating_function}), but with the scattering phase
$\delta_\eta(t)$  turning into the  following sequence of pulses:
\begin{equation}
\label{pulses_series}
\delta_\eta(t)=\sum_{n=0}^\infty\delta_{\eta,n}w_\tau(t,t_n)\,,
\end{equation}
with  partial phase shifts 
\begin{eqnarray}&&
\label{phase_center}
\delta_{\eta,2n}=\eta\lambda t_{-\eta}\sqrt{K} r_\eta^nr_{-\eta}^n
\equiv \eta\lambda e^*_{\eta,2n} \,, \\&&
\delta_{\eta,2n+1}=\eta\lambda t_{-\eta}\sqrt{K}
r_\eta^{n+1}r_{-\eta}^n \equiv \eta\lambda e^*_{\eta,2n+1} \,.
\end{eqnarray}
The time moment $t_n=(n+1/2-1/2K)L/u$ corresponds to the beginning of
the $n$-th pulse.

The result is non-trivial when the measurement interval $\tau$ is
small compared to the length of the LL conductor, $\tau\ll L/u$. (In the 
opposite   limit the pulses overlap and one recovers the results of
the FCS of non-interacting fermions \cite{Levitov-noise}.) 
As for the tunneling spectroscopy problem, the determinants then
split into a product of determinants for individual pulses. 
The FCS of the LL then becomes a superposition of FCS of
non-interacting electrons with fractional charges $e^*_{\eta,n}$.  
For the case of smooth
boundaries we obtain  only one fractional charge, $e^*_{\eta,0}=
\sqrt{K}$ (Table~\ref{table1}). 
In the opposite limit of sharp boundaries, we obtain a
sequence of fractional charges of the form $e^*_{\eta,n}=
2K(1-K)^n/(1+K)^{n+1}$.

These results describe current fluctuations in the interacting part of
the wire. The FCS measured in the non-interacting part (keeping the
assumption $\tau \ll L/u$) has a similar structure, but the values of the 
fractional charges are different, see Ref.~\onlinecite{GGM_short2010}.

\subsection{Luttinger liquid of bosons: Tonks-Grirardeau gas}
\label{bosonic_problems}

We now consider the  problem of a  non-equilibrium LL formed by bosons
with strong repulsion. Interacting bosons out of equilibrium attract currently a
great deal of attention, in particular, in connection with experiments
on cold atoms \cite{cold_atoms}.  

To make the connection between the fermionic and the 
bosonic LL particularly transparent, we will adopt a toy model with
the same setup,  and with  the same behavior of the LL interaction constant
$K(x)$ as the one assumed for the fermionic models considered above. 
 Specifically, we will assume that $K(x)$ takes takes the  value 1 in the
reservoirs, and a value $K$ in the central part of the setup. For 
a gas of fermions 
 $K=1$ implies  the absence of interaction; now, in the present context,   
 it corresponds to hard-core bosons (whose many-body wave function
vanishes when the coordinates  of two particles  coincide); this  system is 
known as the Tonks-Girardeau gas\cite{Girardeau}. The wave function of
this system is 
related to that of non-interacting fermions via the transformation 
\begin{equation}
\label{bosons-to-fermions-1}
\psi_B(x_1,\ldots,x_N)= s(x_1,\ldots,x_N) \psi_F(x_1,\ldots,x_N)\,, 
\end{equation}
where $s(x_1,\ldots,x_N)$ is a sign factor ($\pm 1$) counting the
parity of the number of permutations of  coordinates,
\begin{equation}
\label{bosons-to-fermions-2}
s(x_1,\ldots,x_N) = \prod_{i>j}^N \sgn(x_i-x_j)\,.
\end{equation}
If the fermionic wave function is real, this simply yields  $\psi_B =
|\psi_F|$. 
We will further assume that after this boson-to-fermion
transformation, the fermions in the reservoirs are characterized by
distribution functions $n_\eta(\epsilon)$, as in the above fermionic
models. While very natural in the fermionic language, this requirement
is in general quite artificial for bosons. We do not know whether it
can be realized in an experiment and consider this as a theoretical
toy model. A more realistic situation arises in the case of partial
non-equilibrium, where each of the reservoirs is at equilibrium 
but the temperatures of the two reservoirs  are different. 

We will analyze the single-particle bosonic Green functions
\begin{eqnarray}&&
G_B^>(x,t)=-i\langle\Psi_B(x,t)\Psi^\dagger_B(0,0)\rangle\, \,\,,  \\&&
G_B^<(x,t)=-i\langle\Psi^\dagger_B(0,0)\Psi_B(x,t)\rangle \,\,\,  .\\&&
\end{eqnarray}
that carry information about spectral properties (density of states
and distribution functions) of the system. 
To calculate them, we proceed, in analogy with fermionic systems, via
the non-equilibrium bosonization technique. The term ``bosonization'' here is,
perhaps, not optimal since the original system is
bosonic to begin with. What actually happens is a transformation from the
original bosonic fields $\Psi_B$ to new bosonic fields $\phi, \theta$,
the latter  describing
density fluctuations in the system. The original field operator is
expressed in term of the new fields as 
\cite{Haldane}  
\begin{equation}
\label{bose_operator}
\Psi^\dagger_B(x)=\sqrt{\rho_0+\Pi(x)}\bigg\{\sum_{m \in {\rm even}
}e^{im\phi(x)}{\bigg\}}e^{-i\theta(x)}\,. 
\end{equation}
Here the field $\phi(x)$ is related to the smeared density $\rho(x) = \rho_0 +
\Pi(x)$ (where $\rho_0$ is the average density) via $\rho(x)
=-\partial_x\phi(x)/\pi$. 
The  bosonic fields ($\phi, \theta$) satisfy the commutation relation
\begin{equation}
[\phi(x),\theta(x')]=\frac{i\pi}{2}\rm{sgn}(x-x')\,\,.
\end{equation}
The bosonization  prescription (\ref{bose_operator}) may be further
simplified by discarding fast oscillating terms (corresponding to
$m\ne 0$ in Eq.~(\ref{bose_operator})) and neglecting $\Pi(x)$ in
comparison to $\rho_0$ in the preexponential factor. We thus obtain
\begin{equation}
\Psi^\dagger_B(x)\simeq\sqrt{\rho_0}e^{-i\theta(x)}\,.
\end{equation}
The goal of the discussion above was to present the correlators 
needed for $G_B^\gtrless(x,t)$ in terms of the bosonic fields
$\phi, \theta$. We next discuss the pertinent action. 

Since the boson-to-fermion transformation (\ref{bosons-to-fermions-1})
preserves the density $|\psi|^2$, and in view of our assumption about
the density matrix in the reservoirs, the Keldysh action of the
non-equilibrium Tonks-Girardeau gas has exactly the same form as for
the corresponding problem of free fermions \cite{GGM_long2010}, 
\begin{equation}
S_0[\rho,\bar{\rho}]=\sum_\eta S_0[\rho_\eta,\bar{\rho}_\eta]\,\,, 
\end{equation}
where $\rho$ and $\bar{\rho}$ are the classical and quantum components
in Keldysh representation, 
and the action for excitations with a  given chirality is 
\begin{equation}
S_0[\rho_\eta,\bar{\rho}_\eta]=-\rho_\eta\Pi_\eta^{a^{-1}}\bar{\rho}_\eta-i\ln
Z_\eta[\bar{\chi}_\eta]\,. 
\end{equation}
The information about the non-equilibrium state of the problem is encoded
in the infinite sum of vacuum loops 
\begin{equation}
i\ln Z_\eta[\bar{\chi_\eta}]
=\sum_{n=2}^\infty \frac{i^{n+1}}{n!}\bar{\chi}_\eta^nS_{n,\eta}\,,
\end{equation}
representing a partition function of free fermions subject to the
external quantum field
\begin{equation}
\bar{\chi}_\eta=\Pi_\eta^{a^{-1}}\bar{\rho}_\eta\,.
\end{equation}
Here $S_{n,\eta}$ is the $n$-th order density cumulant of free
fermions in the given non-equilibrium state . 
The right and left density components are related to the fields $\phi$
and $\theta$ via
\begin{equation}
\rho_\eta(x)=\frac{\eta}{2\pi}\partial_x\phi_\eta\,,
\end{equation}
where
\begin{eqnarray}&&
\phi_R=\theta -\phi\,, \nonumber  \\&& 
\phi_L=\theta+\phi \,.
\end{eqnarray}
In order to find the Green functions of the original bosons  one therefore needs
to calculate the  functional integral  
\begin{eqnarray}
\label{hard_core_bosons_funct_int}&&
G_B^>(x,t)=-i\rho_0\int {\cal D}\phi e^{iS[\rho,\bar{\rho}]}\nonumber \\&& 
\times
e^{(-i/\sqrt{2})[\theta[0,0]-\theta[x,t]+\bar{\theta}(0,0)+\bar{\theta}(x,t)]}\,\,, 
\end{eqnarray}
and similarly for $G^<$.
This   functional integral is fully analogous to the one we had to
evaluate while studying the Green functions of free fermions 
(see  Ref.~\onlinecite{GGM_long2010} for details).
The only  difference is  in the source part,
 i.e., in  the exponent that contains terms linear in bosonic fields. 
This difference has a rather transparent meaning: creation of the
fermion $\psi_\eta$  
corresponds to the vertex operator  $\Psi_{F\eta}^\dagger \propto e^{-i\phi_\eta}$ 
that generates a soliton (a step-like plasmon wave with amplitude
$2\pi$) that propagates in the  direction $\eta$.   
By contrast, bosons represented by the  operators $\Psi_B$ do
not have chirality; the corresponding creation operator 
$\Psi_B^\dagger
\propto e^{-i\theta} = e^{-{i\over 2}(\phi_R + \phi_L)}$ generates both 
right and left moving waves. This results in  two
Fredholm determinants---one corresponding to the left and another to
the right reservoir---both  with  scattering phases  $\pi$ 
(as opposed to a single determinant with a scattering phase $2\pi$ for
free fermions):  
\begin{eqnarray}&&
G^>_B(x,\tau)=-i\rho_0\Delta_{R,\tau-x/v}(\pi)\Delta_{L,\tau+x/v}(\pi)\nonumber
\\&& 
\times e^{-i\frac{\pi\tau}{8}[\sgn(\tau+x/v)+\sgn(\tau-x/v)]}\,.
\label{G-Tonks-Girardeau}
\end{eqnarray}
Here the numerical prefactor has been fixed by comparison with the
equilibrium case, see Appendix \ref{Tonks_girardo_equilibrium}.
At zero temperature $\Delta_\tau(\delta) \sim
(1+\Lambda^2\tau^2)^{-{1\over2}(\delta/2\pi)^2}$, which reproduces the
  well-known $x^{-1/2}$ (for $\tau=0$) behavior corresponding to the
  $k^{-1/2}$ momentum distribution of particles in the equilibrium
  $T=0$ Tonks-Girardeau gas.

Equation (\ref{G-Tonks-Girardeau}) yields the Green function of the
non-equilibrium Tonks-Girardeau gas with a  LL constant $K=1$. We now
assume that the interaction is different in the central part of the
system, so that $K(x)$ there is equal to $K\ne 1$. Performing the
analysis in analogy with the fermionic problem, we find
\begin{equation}
G^>_B(x,\tau)=-i\rho_0\Delta_R[\delta_R(t)]\Delta_L[\delta_L(t)]
e^{-i\frac{\pi\tau}{4}\sgn(\tau)}
\,,
\label{G-Tonks-Girardeau-b}
\end{equation}
where each of the phases $\delta_\eta(t)$ is given by the arithmetic
mean of the corresponding phase for fermionic Green functions $G_{FR}$
and $G_{FL}$. In general we obtain again an infinite sequence of pulses,
as for the fermionic problem. In the case of  smoothly varying $K(x)$
only the first pulse survives  in each of the phases $\delta_\eta(t)$; 
its amplitude is equal to $\pi/\sqrt{K}$, as shown in Table~\ref{table1}.

\subsection{Summary}

Let us briefly summarize the results presented
above. All the problems considered (Fermi edge singularity,
tunneling spectroscopy of interacting fermions, their full counting
statistics, and spectral properties of interacting bosons) can be
solved by the non-equilibrium bosonization approach, with  the results
expressed in terms of Fredholm determinants of counting
operators. In these expressions, 
all differences between the problems are encoded in the
values of scattering phases $\delta_\eta(t)$. 
These scattering phases consist either of one pulse (for the FES
problem and LL problem with smooth boundaries), or of a sequence of
well-separated pulses (for a sufficiently long LL sample with  sharp
boundaries). In the latter case the determinant can be split into  a
product of determinants, each of which corresponding to a single pulse.
Therefore, physical properties of a number of many-body problems are
governed by the behavior of Fredholm determinants with a single phase pulse.
The analysis of this behavior is presented in the next section.

It is worth mentioning that there is a vast literature on the
connection of Fredholm determinants to counting statistics as well as to
quantum and classical integrable models and free-fermion problems; see,
in particular,  
Refs.~\onlinecite{Levitov-noise,Klich,Muzykantskii03,Shelankov03,
Braunecker06,Schoenhammer07,Avron08,Hassler08,Abanov09,Jimbo80,Izergin98,
bettelheim06,Zvonarev,Nazarov,Marquardt}.

\section{Asymptotic properties of Fredholm determinants} 
\label{Fredholm_determinant}

\subsection{Ultraviolet regularization and reduction to Toeplitz form}
\label{s:Toeplitz}

We consider a Fredholm determinant of the counting operator for
scattering  phase exhibiting  a single pulse of an amplitude $\delta$
and duration $\tau$, 
\begin{equation}
\label{single_pulse}
 \delta(t)=\delta \times w_\tau(t,0)\,.
\end{equation}
In this case the Fredholm determinant (\ref{determinant_definition})
is of the Toeplitz type.
To show this, one  defines a projection operator  $\hat{P}$ that acts on a
function $y(t)$ by restricting it to  the time interval $[0,\tau]$:
\begin{eqnarray}
\hat{P} y(t)=\left\{
\begin{array}{l}
      y(t)\, , \,\,\, {\rm for}\,\,\,  t \in [0,\tau] 
      \\ \\
      0 \,, \,\,\,\,\,\,\,\,\,\,\,  {\rm  otherwise}\,.
\end{array}
\right.
\label{projection_operator}
\end{eqnarray}
For  the single-pulse scattering phase (\ref{single_pulse}) 
Eq.~(\ref{determinant_definition}) can then be rewritten in the form 
\begin{equation}
\label{determinant_definition2}
\Delta[\delta(t)]=
\det[1 + \hat{P}(e^{-i\delta}-1) \hat{n}]\,\,.
\end{equation}
Since  $\hat{P}^2=\hat{P}$, we can bring  the 
determinant (\ref{determinant_definition2})  to the form
\begin{equation}
\label{determinant_Toeplitz}
\Delta[\delta(t)]=
\det[1
  + \hat{P}(e^{-i\delta}-1) n \hat{P}]\,\,.
\end{equation}
This determinant still requires an ultraviolet regularization. A
possible way to introduce it is to discretize the coordinate $t$ by
introducing an elementary unit of the size   $\Delta t=\pi/\Lambda$, 
such that $t_j=j \Delta t$. This corresponds to restricting the energy
$\epsilon$ variable to the range $[-\Lambda,\Lambda]$. 
In the formulations of Szeg\H{o} and Hartwig-Fisher theorems that will be
applied below the function $f(z)$,
that generates the matrix is defined in the complex plane of variable $z$, 
and its values on the unit circle $z=e^{i\theta}$ 
parametrized by the angle $\theta\in [-\pi,\pi]$ are
important. In Eq.~(\ref{determinant_Toeplitz}) this function is
$f(\epsilon) = 1 + n(\epsilon) (e^{-i\delta} -1)$.  
The correspondence between  $\epsilon$ and $\theta$ is established by rescaling the energy
$\pi\epsilon/\Lambda = \theta$. Further, we need to eliminate the
jump in $n(\theta)$ at $\theta = \pm \pi$ that results from a hard
cutoff and would generate an additional, unphysical contribution of
the Fermi-edge type. This is done by introducing a phase factor,
\begin{equation}
\label{f_epsilon_periodic}
f(\epsilon) = [1 + n(\epsilon) (e^{-i\delta} -1)] e^{-i{\delta\over
    2}{\epsilon\over \Lambda}}\,,
\end{equation}
that makes $f(\epsilon)$ periodic on $[-\Lambda,\Lambda]$. 
Fourier transforming the periodic function  $f(\epsilon)$, we obtain 
$f(t_j)$ with $t_j=j \pi/\Lambda$.  
Equation
(\ref{determinant_Toeplitz})  then reduces to a determinant 
of a large ($N \times N$, where
$N=\tau \Lambda/\pi$) but finite matrix   
\begin{equation}
\label{Toeplitz}
\Delta_N[f]=\det[f(t_j-t_k)]\,, \,\,\,  0\leq j,k \leq N-1\,.
\end{equation}
In the above derivation we assumed $\tau>0$; the result for $\tau<0$
follows from  the property  
$\Delta_{-\tau}(\delta)=\Delta_\tau(-\delta)$.

The matrix $\{f(t_j-t_k)\}$ with $0\leq j,k \leq N-1$ is of a Toeplitz
form. Bellow this will allow us to apply  known mathematical results 
concerning the asymptotic properties of its determinant $\Delta_N$ 
in the limit of large $N$. 
Physically, this corresponds to the regime of long time
$\tau$, i.e. to infrared asymptotics of correlation functions under
interest. For arbitrary times $\tau$ 
Eqs.~(\ref{Toeplitz}), (\ref{f_epsilon_periodic}) can be directly used
for numerical evaluation of the determinant $\Delta_N[f]$. 

The following important point is also worth stressing. The Green functions
of LL tunneling spectroscopy and FES problems require
evaluation of such determinants $\Delta(\delta)$ at phases $\delta$ that are not
small. For example, in the case of relatively weak LL interaction ($K$
close to unity) or small phase shift $\delta_0$ for the scattering on
the core hole in the FES problem, one needs to know the determinant
at $\delta$ close to $2\pi$, see Table~\ref{table1}. For strong LL
interaction or large phase shift $\delta_0$ the value of $\delta$ can
be, in principle, arbitrarily large. As was discussed in our papers
\cite{GGM_long2010} the determinant $\Delta(\delta)$ should be then
understood as analytically continued from the region of small
$\delta$. We emphasize now that the present regularization
(discretization of time and introduction of the phase factor ensuring
periodicity in energy) implements the required analytic continuation. 
Indeed, 
in the original form of the determinant, Eq.~(\ref{determinant_definition2}),
the information about the integer part of $\delta/2\pi$ was not
explicit, which made the analytic continuation necessary. On the other
hand, in the present regularization the integer part of $\delta/2\pi$ enters
explicitly via the last phase factor in
Eq.~(\ref{f_epsilon_periodic}). As we will demonstrate below, this
allows one to directly compute the determinant at arbitrary large $\delta$.

\subsection{Simplified analysis of asymptotics via Szeg\H{o} formula }
\label{s:Szego}

The long time behavior of determinants of Toeplitz matrices can be
found using  the Szeg\H{o} theorem and its extension known as
Fisher-Hartwig conjecture. The condition of applicability of Szeg\H{o}
theorem requires that $f(z)$ is a sufficiently smooth function. This
condition is not fulfilled in our case. Indeed, already at equilibrium (and at
zero temperature) $f(\epsilon)$ has a jump at the Fermi energy. In
non-equilibrium situations we are interested in $f(\epsilon)$ will have
two (``double-step distribution'') or more such jumps. We will see,
however, that the Szeg\H{o} formula nevertheless yields correctly the 
main ingredients of the result (dephasing rate and modified power-law
exponents). A more accurate treatment will be performed below in
Sec.~\ref{s:Fisher-Hartwig} in the framework of Fisher-Hartwig
formula.

Here we assume $|\delta|<\pi$. (The accurate consideration in
Sec.~\ref{s:Fisher-Hartwig} will be performed for arbitrary $\delta$.) 
We first consider a simple case  of thermal equilibrium,  
when the  determinant (\ref{determinant_Toeplitz}) can be
calculated explicitly 
\begin{equation}
\Delta_{\tau}(\delta)\simeq\frac{1}{
\left(1+\Lambda^2\tau^2\right)
^{\frac{1}{2}\left(\frac{\delta}{2\pi}\right)^2}
}
\left(
\frac{\pi T \tau}{\sinh \pi T \tau}
\right)^{\left(\delta/2\pi\right)^2}\,.
\label{det_eq}
\end{equation}
Note that the precise behavior of the functional determinant at the
ultraviolet scale, $\tau \sim \Lambda^{-1}$, depends on the 
regularization procedure.
Equation (\ref{det_eq}) corresponds to  a smooth cut-off
$e^{-|\epsilon|/\Lambda}$ in the energy space\cite{GGM_long2010} that
is different from the regularization we use in this work.
We are interested, however, in energy scales much less than
$\Lambda$. i.e. $\tau \gg \Lambda^{-1}$, where the determinant does
not depend on the regularization scheme, up to an overall prefactor
independent on the distribution function $n(\epsilon)$.
At  $T=0$ one readily finds from Eq.~(\ref{det_eq})
\begin{equation}
\label{eq_a1}
\Delta_\tau(\delta) \simeq
(\tau\Lambda)^{-\left(\delta/2\pi\right)^2}\, ,\,\,  \tau \gg
\Lambda^{-1}\,. 
\end{equation}

To apply the Szeg\H{o} formula (see Appendix
\ref{Fisher-Hartwig-conjecture}), we have to calculate the Fourier
transform $V(t_j)$ of $\ln f(\epsilon)$, 
\begin{eqnarray}&&
\label{eq_d1}
V(t_j)=\int_{-\Lambda}^{\Lambda} \frac{d\epsilon}{2\pi}e^{-i\epsilon  t_j} 
\ln f(\epsilon)\,.
\end{eqnarray}
where $f(\epsilon)$ is given by Eq.~(\ref{f_epsilon_periodic}). 
For the case of the Fermi-Dirac distribution with $T=0$ one finds
\begin{eqnarray}&&
\label{Veq}
V(t_j)=\frac{\delta}{2\pi} \times
\left\{
\begin{array}{ll}
1/t_j\ , & \qquad t_j \ne 0\ ; \\
-i\Lambda\ , & \qquad t_j = 0\ .
\end{array}
\right.
\end{eqnarray}
According to the (strong) Szeg\H{o} theorem [Eq. (\ref{Szoego})], 
the large-$N$ behavior of
the determinant reads, in the present notations
\begin{equation}
\label{szego-result}
\Delta_N \sim \exp \left\{N\, \Delta t\, V(0) + \Delta
t\,\sum_{j=0}^Nt_jV(t_j)V(-t_j)\right\}\,.
\end{equation}
The first term in the exponent $N\, \Delta t\, V(0) = \tau V(0) = 
-i\delta\tau\Lambda/2\pi$ is purely imaginary and yields just a phase
factor. Calculating the second term, we find  
\begin{eqnarray}&&
\label{eq_b1}
\Delta t \sum_{j=0}^N t_j V(t_j) V(-t_j) \simeq
-\left(\frac{\delta}{2\pi}\right)^2\int_{\Lambda^{-1}}^\tau\frac{dt}{t}=\nonumber
\\&& 
=-\left(\frac{\delta}{2\pi}\right)^2\ln (\tau\Lambda)\,.
\end{eqnarray}
Here we assumed that the time is sufficiently long (compared to the
ultraviolet scale),  $\tau\Lambda \gg 1$, which is
exactly the condition of applicability of the Szeg\H{o} theorem ($N\gg
1$). 
Exponentiation of Eq.~(\ref{eq_b1}) according to
Eq.~(\ref{szego-result}) reproduces the result
(\ref{eq_a1}). 

It is easy to verify that the Szeg\H{o} formula yields the correct
behavior of the determinant also at finite temperature $T$. 
(In fact, in the long-time regime $\tau T \gg 1$ the Szeg\H{o} theorem
becomes rigorously applicable.)

We turn now to the non-equilibrium situation and  focus on a double step
distribution function,  
\begin{equation}
\label{double-step}
n(\epsilon)=(1-a)n_0(\epsilon-\epsilon_0) + an_0(\epsilon-\epsilon_1)\,.
\end{equation}
Here  $\epsilon_0= -aU$, $\epsilon_1= (1-a)U$, and
$n_0(\epsilon)$ is the zero-temperature Fermi-Dirac function, 
$n_0(\epsilon)=\theta(-\epsilon)$.


The function $f(\epsilon)$ has now the form 
\begin{eqnarray}
\label{eq_a2}
f(\epsilon)&=& e^{-i{\delta\over 2} {\epsilon \over \Lambda}}
\nonumber \\
& \times & \left\{
\begin{array}{ll}
      e^{-i\delta} \,, & \ \   -\Lambda < \epsilon < \epsilon_0
      \\ 
      1+\left(e^{-i\delta}-1\right)a  \,, & \ \ \epsilon_0 < \epsilon
      < \epsilon_1 
\\  
1 \,, & \ \ \epsilon_1 < \epsilon
      < \Lambda\,.
\end{array}
\right.
\end{eqnarray}
The Fourier transform of $\ln f$, Eq.~(\ref{eq_d1}), reads
\begin{equation}
V(t_j)= \left\{
\begin{array}{ll}
-\frac{1}{t_j}\left(\beta_0e^{-i\epsilon_0t}+\beta_1e^{-i\epsilon_1t}\right)
\,, & \ \ t_j\ne 0 \\[0.3cm]
-i{\delta\Lambda\over 2\pi} + iU\left(a{\delta\over 2\pi} +
  \beta_1\right)\,, & \ \ t_j=0\,,
\label{v_t_double_step}
\end{array}
\right.
\end{equation}
where we defined
\begin{eqnarray}&&
\label{eq_a4}
\beta_1=-\frac{i}{2\pi}\ln\bigg[1+\left(e^{-i\delta}-1\right)a\bigg] \,,
\nonumber \\&&
\beta_0=-\frac{\delta}{2\pi}-\beta_1\,.
\end{eqnarray} 
Now we apply the Szeg\H{o} theorem. The leading, linear-in-$t$, term in the
asymptotics of $\ln \Delta$ is governed by $V(0)$ given by the second
line in Eq.~(\ref{v_t_double_step}). Particularly important is the real
part of $V(0)$ that leads to the exponential decay of the determinat
with time, $\Delta \propto \exp( - \tau/2\tau_\phi)$. The
corresponding decay rate is given by 
\begin{equation}
\label{dephasing1}
\tau_\phi^{-1} =2{\rm Re}\: V(0)=
-\frac{U}{2\pi}\ln\bigg[1-4a(1-a)\sin^2\frac{ \delta}{2}\bigg]\,. 
\end{equation}
To find the subleading term in the Szeg\H{o} formula (\ref{szego-result}),
we have to evaluate the sum $\Delta t\sum_{k=1}^N  t_k V(t_k)
V(-t_k)$. We get 
\begin{eqnarray}
&&
\Delta t\sum_{k=1}^N  t_k V(t_k) V(-t_k) \\ \nonumber
&& \simeq 
-\int_{\Lambda^{-1}}^\tau {dt\over t} (\beta_0^2 + \beta_1^2 + 2
\beta_0\beta_1\cos Ut) \\ \nonumber
&& 
\simeq \left\{
\begin{array}{ll}
-\left({\delta\over2\pi}\right)^2 
\ln \Lambda \tau\ , & \ \ \tau \ll U^{-1}\\[0.3cm]
-\left({\delta\over2\pi}\right)^2  \ln {\Lambda\over U} - (\beta_0^2 +
\beta_1^2)\ln U\tau 
\ , & \ \ \tau \gg U^{-1}\,. 
\end{array}
\right.  \nonumber \\
&&
\label{szego-subleading-double-step}
\end{eqnarray}
In the short-time regime, $U\tau\ll 1$, we simply reproduce the
equilibrium result. On the other hand, for long times,  $U\tau\gg 1$,
a different behavior emerges,
\begin{equation}
\label{eq_a9}
\Delta_\tau(\delta) \sim e^{-\tau/2\tau_\phi}
(\Lambda/U)^{-(\delta/2\pi)^2}(U\tau)^{-(\beta_0^2+\beta_1^2)}\,, 
\end{equation}
where  $\beta_0, \beta_1$ are given by Eq.(\ref{eq_a4}) and $\tau_\phi$
by Eq.~(\ref{dephasing1}). 

As has been already mentioned, the present problem goes, strictly
speaking, beyond the range of applicability of the Szeg\H{o} theorem,
since  the function $f(\epsilon)$ has discontinuities. This
results in the (correct) $\ln N$ behavior of the second term in the exponent of
(\ref{szego-result}), while it should have a constant
limit as $N\to \infty$ under the conditions of applicability 
of the Szeg\H{o} theorem. It turns out, however, that the key results
obtained above---dephasing rate and modified power-law exponents---are
correct. This will be shown in Sec.~\ref{s:Fisher-Hartwig} by using
recent mathematical results on Fisher-Hartwig conjecture which treats
Toeplitz determinants of exactly the type we have encountered.
The application of the Fisher-Hartwig formalism will allow us 
not just to confirm the above results but also to go considerably
further. First, we will calculate the asymptotics of the determinants
exactly, including prefactors $\sim N^0$. Second, we will
obtain results for an arbitrary phase $\delta$. As we have already
emphasized, this is important for the analysis of the many-body problems
considered above. Third, we will obtain  not only the leading contribution
but also subleading terms. As we show below, various contributions
have very transparent physical meaning in the problems of FES and
tunneling spectroscopy, corresponding to power-law behavior at
multiple Fermi edges.

\subsection{Accurate analysis of asymptotics via Fisher-Hartwig conjecture}
\label{s:Fisher-Hartwig}

It is instructive to begin again by considering the zero-temperature 
equilibrium case, $n(\epsilon) = \theta(-\epsilon)$.  
In the case of a single singular point, the Fisher-Hartwig 
generating function $f(z)$ has the form
\begin{equation}
\label{a5}
f(z)=e^{V(z)}|z-z_0|^{\alpha_0}
\left(\frac{z}{z_0}\right)^{\beta_0}g_{z_0,\beta_0}(z)\,,
\end{equation}
where $g_{z_0,\beta_0}(z)$ is a ``jump function'',
Eq.~(\ref{jump-function}). 
Comparing Eq.~(\ref{a5}) with  Eq.~(\ref{f_epsilon_periodic})
(where we set  $\theta = \pi \epsilon/\Lambda$ and $z=e^{i\theta}$), 
we identify the parameters:
$z_0=1$, $\alpha_0=0$, $\beta_0=-\delta/2\pi$, and $V(z) = - i\delta/2$. 
Using the Fisher-Hartwig formula (\ref{fisher-hartwig}), we obtain the
asymptotics 
\begin{equation}
\label{fh-equilibrium}
\Delta_N(\delta) =
e^{-i{\delta N\over 2}} N^{-(\delta/2\pi)^2}  
G\left(1-{\delta\over2\pi}\right)G\left(1+{\delta\over2\pi}\right)\,, 
\end{equation}
where $G(z)$ is the Barnes $G$-function.
In physical notations $N=\tau\Lambda/\pi$. 
It is now easy to see that the Fisher-Hartwig formula (\ref{fisher-hartwig})
reproduces correctly the power-law behavior (\ref{eq_a1}) of the
determinant. 

The result (\ref{fh-equilibrium}) is valid for arbitrary
$\delta$. Consider an important case of $\delta$ in the
vicinity of $2\pi$, $\delta=2\pi+\delta'$ with $|\delta'|\ll 1$, 
that is relevant to the FES problem with small
scattering phase and to tunneling spectroscopy of LL with weak
interaction. Using $G(1) = 1$ and $G(z) \simeq z$ for small $z$, we get
\begin{equation}
\label{fh-equilibrium-near-2pi}
\Delta_\tau(\delta) \simeq - {\delta'\over 2\pi} 
e^{-i\tau\Lambda(1+\delta'/2\pi)}
\left({\tau\Lambda\over\pi}\right)^{-(1+\delta'/2\pi)^2}\,.
\end{equation}
It was shown earlier \cite{GGM_long2010} that
$\Delta_\tau(\delta)$ at $\delta\to 2\pi$ should yield, up to a
proportionality factor,
the free fermion Green function $G_0(\tau)$, see Eq.~(\ref{a1}). 
As we see from 
Eq.~(\ref{fh-equilibrium}), the exact correspondence in
the present ultraviolet regularization of the determinant is 
\begin{equation}
\label{fh-equilibrium-free-fermions}
G_0^\gtrless(\tau) = e^{i\tau\Lambda}{\Lambda\over \pi v} 
{1\over 1\mp i/\Lambda\tau} \: \lim_{\delta'\to 0}
{1\over \delta'} \Delta_\tau(2\pi+\delta')\,.
\end{equation}
We will demonstrate below how this general formula works for the case
of a multiple-step distribution. 

We now turn to the case of a double-step distribution
(\ref{double-step}). The singular points are
$z_j=e^{i\pi\epsilon_j/\Lambda}$ ($j=0,1$), where $\epsilon_0 = -aU$,
$\epsilon_1 = \epsilon_0 + U$. We consider first the case of
$|\delta|<\pi$. 
It is easy to see that Eq.~(\ref{eq_a2}) is of Fisher-Hartwig form of
generating function with two jump-type singularities, 
\begin{equation}
\label{eq_a3}
f(z)=e^{V(z)}\left(\frac{z}{z_0}\right)^{\beta_0}
\left(\frac{z}{z_1}\right)^{\beta_1}
g_{z_0,\beta_0}(z)g_{z_1,\beta_1}(z)\,, 
\end{equation}
with $\beta_j$ ($j=0,1$) given by Eq.~(\ref{eq_a4}) and  
\begin{equation}
\label{V_FH_double_step}
V(z) = {\rm const} = {-i\delta\over 2} + i {U\pi\over\Lambda}\left({a\delta\over
    2\pi} + \beta_1\right)\,.
\end{equation}
The logarithm in Eq.~(\ref{eq_a4}) and in analogous formulas for
$\beta_j'$ below is understood in the sense of its main
branch (with imaginary part between $-\pi$ and $\pi$).
We further note that under the condition
$U \ll \Lambda$ we can approximate $|z_1-z_0| \simeq
\pi U/\Lambda$. Applying Eq.~(\ref{fisher-hartwig}), we thus get
\begin{eqnarray}
\label{fh4}
\Delta_\tau(\delta) &\simeq & \exp\{- {i\delta\over 2\pi}\tau\Lambda
-i\tau\mu - \tau/2\tau_\phi \} \nonumber \\
& \times & \left( {\tau\Lambda\over\pi}\right)^{-\beta_0^2-\beta_1^2}
\left( {\Lambda\over\pi U}\right)^{-2\beta_0\beta_1} \nonumber \\
& \times &
G(1+\beta_0)G(1-\beta_0)G(1+\beta_1)G(1-\beta_1)\,, \nonumber \\
&&
\end{eqnarray}
 where  $1/\tau_\phi$ is the exponential decay (dephasing) rate,
 Eq.~(\ref{dephasing1}), 
and $\mu = -U({\rm Re} \beta_1 + a\delta/2\pi)$.
This confirms the  long-time ($U\tau\gg 1$) behavior of the
determinant obtained above from the Szeg\H{o} formula, 
Eq.~(\ref{eq_a9}), and yields the exact value of the corresponding
prefactor. 

Consider now the more general situation, when $\delta$ is not small:
$\delta = 2\pi M + \delta'$ with some integer $M$ and
$|\delta'|<\pi$. 
The exponents $\beta_0$ and $\beta_1$ can be now chosen to be 
\begin{eqnarray}
\beta_1 &=& - {i\over 2\pi} \ln(1-a+a e^{-i\delta}) \equiv \beta_1'\,,
\label{fh1} \\
\beta_0 &=& - {\delta\over 2\pi} - \beta_1 = -M - {\delta'\over 2\pi}
- \beta_1' \equiv - M  + \beta_0'\,.
\label{fh2}
\end{eqnarray}
We have introduced here $\beta_0'$ and $\beta_1'$ satisfying
$|{\rm Re} \beta_j'|<1/2$, $\beta_0'+\beta_1' = -\delta'/2\pi$.  The exponents
$\beta_j$ (that may differ from $\beta'_j$ by an integer only) satisfy
$\beta_0+\beta_1 = -\delta/2\pi \equiv -M -\delta'/2\pi$. Equations
(\ref{fh1}), (\ref{fh2}) represent 
one possible choice; the final result will not depend
on a particular choice in view of the summation over integers $n_j$ in
Eq.~(\ref{fisher-hartwig-general}).  
We obtain from Eq.~(\ref{fisher-hartwig-general}):
\begin{eqnarray}
\label{fh5}
\Delta_\tau(\delta) &\simeq & \exp\{- {i\delta\over 2\pi}\tau\Lambda
-i\tau\mu' -iM\epsilon_0\tau - \tau/2\tau_\phi \} \nonumber \\
& \times & \sum_{n=-\infty}^\infty
\left( {\tau\Lambda\over\pi}\right)^{-(\beta'_0-M+n)^2-(\beta'_1-n)^2} \nonumber\\
&\times&
\left( {\Lambda\over\pi U}\right)^{-2(\beta_0'-M+n)(\beta_1'-n)} \nonumber \\
& \times & e^{-inU\tau}
G(1+\beta_0'-M+n)G(1-\beta_0'+M-n)\nonumber\\
&\times& G(1+\beta_1'-n)G(1-\beta_1'+n)\,, \nonumber \\
&&
\end{eqnarray}
where  $\mu' = -U({\rm Re}\: \beta'_1 + a\delta'/2\pi)$. When $M=0$,
the dominant term (that is characterized by the smallest power-law exponent 
${\rm Re}[(\beta'_0-M+n)^2+(\beta'_1-n)^2]$) is the one with $n=0$, 
reproducing Eq.~(\ref{fh4}).

In the
particularly interesting case of $M=1$, when $\delta$ is in the
vicinity of $2\pi$, the two leading terms are those with $n=0$ and
$n=1$. Retaining only these terms, we find
\begin{eqnarray}
\label{fh6}
\Delta_\tau(\delta) &\simeq & \exp\{- {i\delta\over 2\pi}\tau\Lambda
-i\tau\mu' - \tau/2\tau_\phi \} \nonumber \\
& \times & 
\left( {\tau\Lambda\over\pi}\right)^{-(\beta'_0)^2-(\beta'_1)^2}
\left( {\Lambda\over\pi U}\right)^{-2\beta_0'\beta_1'} 
\left(-{\delta'\over 2\pi}\right)\nonumber \\
& \times & \left[ (1-a)e^{iaU\tau}\left( {\tau\Lambda\over\pi}\right)^{2\beta_0'}
\left( {\Lambda\over\pi U}\right)^{2\beta_1'} \right. \nonumber \\
&+& \left. a e^{i(a-1)U\tau}\left( {\tau\Lambda\over\pi}\right)^{2\beta_1'}
\left( {\Lambda\over\pi U}\right)^{2\beta_0'}\right]\,.
\end{eqnarray}

In the limit $\delta'\to 0$, substituting Eq.~(\ref{fh6}) into
Eq.~(\ref{fh-equilibrium-free-fermions}), we correctly reproduce the
free-electron Green function for the double-step distribution,
$G^\gtrless_0(\tau)= [(1-a)e^{iaU\tau} + a
e^{(1-a)U\tau}]G^\gtrless_{0,T=0}(\tau)$, where
$G^\gtrless_{0,T=0}(\tau)$ is the equilibrium, $T=0$ value of
$G^\gtrless_0(\tau)$. 

\begin{figure}
\includegraphics[width=1\columnwidth,angle=0]{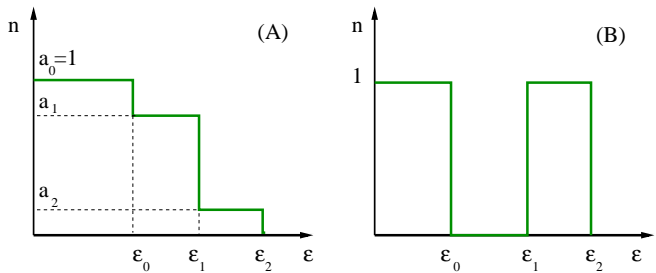}
\caption{
Distribution functions with multiple edges. Two
examples of three-step distributions are shown:
(A) with monotonously decreasing occupation;
(B) with population inversion.}
\label{figure1}
\end{figure}

These results can be generalized to a multi-step
distribution, Fig.~\ref{figure1}. 
Consider a distribution function of the form
\begin{equation}
\label{multistep-distribution}
n(\epsilon) = \left\{
\begin{array}{ll}
1 \equiv a_0 \,, & \qquad \epsilon < \epsilon_0 \\
a_1\,, & \qquad \epsilon_0 < \epsilon <\epsilon_1 \\
\ldots &  \\
 a_m\,, & \qquad \epsilon_{m-1} < \epsilon <\epsilon_m \\
0 \equiv a_{m+1}\,, & \qquad \epsilon_m < \epsilon\,.  
\end{array}
\right.
\end{equation}
Here all $a_j$ with $j=1,\ldots,m$ satisfy $0\le a_j \le 1$ without any further
restrictions. In particular, no requirement of monotonicity is
imposed: the distribution  $n(\epsilon)$ can describe inversion of
population in some regions of energy, with $a_{j+1} > a_j$.
Using Eq.~(\ref{fisher-hartwig-general}), we obtain for $\delta = 2\pi
M +\delta'$ where, as before, $M$ is the integer closest to $\delta/2\pi$:
\begin{eqnarray}
\label{fh7}
\Delta_\tau(\delta) &\simeq & \exp\{- {i\delta\over 2\pi}\tau\Lambda
-i\tau\mu' - \tau/2\tau_\phi \} \nonumber \\
& \times & \sum_{n_0+\ldots+n_m=-M} \exp\{i\tau\sum_jn_j\epsilon_j\}
\left( {\tau\Lambda\over\pi}\right)^{-\sum_j\beta_j^2} \nonumber\\
&\times&
\prod_{j<k}\left( {\Lambda\over\pi U_{jk}}\right)^{-2\beta_j\beta_k} \nonumber \\
& \times & \left. \prod_j
G(1+\beta_j)G(1-\beta_j)\right|_{\beta_j = \beta_j'+ n_j}\,.
\end{eqnarray} 
Here the exponents $\beta_j'$ (satisfying $|{\rm Re} \beta_j'|<1/2$) are
\begin{equation}
\label{fh8}
\beta_i' = -{i\over 2\pi}\left[\ln(1-a_i+a_ie^{-i\delta}) -
\ln(1-a_{i+1}+a_{i+1}e^{-i\delta})\right]\,, 
\end{equation} 
the dephasing rate reads
\begin{eqnarray}
\label{fh9}
{1\over \tau_\phi} &=& 2{\rm Im}\sum_j \beta_j'\epsilon_j
\nonumber \\ &=& - {1\over
  2\pi} \sum_{j=1}^m (\epsilon_j-\epsilon_{j-1})\ln [1 - 4 a_j
  (1-a_j)\sin^2{\delta\over 2}]\,,  \nonumber \\
&&
\end{eqnarray} 
$U_{jk} = |\epsilon_j-\epsilon_k|$,  
and $\mu' = - {\rm Re} \sum_j\beta'_j\epsilon_j$.

The remarkable periodicity in the dependence of the dephasing rate for a
multi-step 
distribution, as a function on the phase $\delta$, should be emphasized.  
When applied to LL spectroscopy, this results in periodic dependence
of $1/\tau_\phi$ on the interaction
strength\cite{GGM_long2010}. The dephasing rate manifests itself in
a broadening of the singularities in the energy space (see
Sec.~\ref{s:energy_representation}), as well as in an exponential
damping of Aharonov-Bohm oscillations in out-of-equilibrium 
interferometry\cite{GGM_long2010}.

While the above results are obtained in time representation, the
experimental measurements of Green functions describing the FES,
Eq.~(\ref{b1}) and the  LL tunneling spectroscopy, Eq.~(\ref{d3}), are
normally performed in energy space. It is thus important to see what
the implications of the above findings in the energy representation are.

\subsection{Singularities in energy representation}
\label{s:energy_representation}

After Fourier transforming  from time into  energy space,
Eq.~(\ref{fh7}) yields multiple power-law singularities of the type
$|\epsilon + \sum_j n_j\epsilon_j-\mu'|^{-1+\sum_j\beta_j^2}$, where
$\beta_j = \beta'_j+n_j$.  
Positions of the singularities are given by linear combinations of
the singular points $\epsilon_j$ of the distribution function, with a
small overall shift $\mu'$. All singularities are broadened by the
dephasing rate $1/2\tau_{\phi}$. 

As an important example, consider the case $M=1$, where $\delta$ is
close to $2\pi$. The leading singularities then correspond to all
$n_j$ being equal to zero except for one, $n_k= -1$. The position of
each such singularity  is close to $\epsilon = \epsilon_k$ and the exponent is
$\sum_j \beta_j^2 = 1 - 2\beta_k' + \sum_j (\beta_j')^2$. There is such
a singularity for each of the singular points of the original
distribution function (i.e. for each $k=0,1,\ldots,m$) as
expected. Further singularities are much weaker. The next ones correspond
to all $n_j$ being zero except for $n_k=n_l=-1$ and $n_p=1$. Such a
singularity is located at $\epsilon = \epsilon_k + \epsilon_l -
\epsilon_p$, with
an exponent $\sum_j \beta_j^2 = 3 - 2\beta_k'- 2\beta_l'+ 2\beta_p' 
+ \sum_j (\beta_j')^2$. The next ones are generated by $n_k=-2$, $n_l=1$,
located at $\epsilon = 2\epsilon_k - \epsilon_l$ and are characterized
by exponents   $\sum_j \beta_j^2 = 5 - 4\beta_k'+ 2\beta_l'+ \sum_j
(\beta_j')^2$, and so on. These subleading singularities can be
understood as resulting from inelastic processes. For example, the
edge $\epsilon = \epsilon_k + \epsilon_l - \epsilon_p$ results from a
creation  of an electron near $\epsilon_k$ accompanied by creation of
a particle-hole pair with energies near $\epsilon_l$ and $\epsilon_p$,
respectively. 

Let us present results for dominant singularities for the FES
Green function (\ref{b1}) in the energy space in an explicit form.  
(The LL tunneling spectroscopy Green function is analyzed in the same way.)
The behavior of the Green functions $G^\gtrless(\epsilon)$ 
for energies close to singular points $\epsilon_k+\mu'$ 
is given by (up to an additive contribution that can be considered as
constant near  $\epsilon_k+\mu'$)
\begin{eqnarray}
\label{fh10}
G^\gtrless(\epsilon) &\simeq& \pm\frac{i}{2v}\prod_{j\neq k}
\left(\frac{\Lambda}{\pi U_{jk}}\right)^{2\beta_j'}
\prod_{j<l}
\left(\frac{\Lambda}{\pi U_{jl}}\right)^{-2\beta_j'\beta_l'}
\nonumber \\
&\times&
(\epsilon-\epsilon_k-\mu'-i/2\tau_\phi)^{\gamma_k} 
\nonumber \\
&\times& \left\{
\begin{array}{ll}
1\mp (a_k-a_{k+1}) \,, & \qquad \epsilon < \epsilon_k+\mu' \\
1\pm (a_k-a_{k+1})\,, & \qquad \epsilon_k+\mu' < \epsilon\,,  
\end{array}
\right.
\end{eqnarray}
where the exponents $\gamma_k$ are given by 
$\gamma_k  = - 2\beta_k' + \sum_j(\beta_j')^2$.

Let us assume for simplicity that all distances between consecutive
singular points are of the same order, $U_{j,j-1} \sim U$. The region
of validity of the behavior (\ref{fh10}) is then
$|\epsilon-\epsilon_k-\mu'|\lesssim U$. Let us emphasize that the
power-law singularity in Eq.~(\ref{fh10}) is smeared by the dephasing
$1/2\tau_\phi$. In a generic situation, when
the phase $\delta' = \delta-2\pi$ is of order $\pi$ (i.e. not small), 
the dephasing rate is of order $U$. Then the smearing is strong, and
the power law essentially does not have room to develop. 
On the other hand, when the phase $\delta = 2\pi + \delta'$ 
is close to $2\pi$ (which
corresponds to a weak interaction in LL or to small phase shift
for scattering on core hole in the FES problem), the
dephasing rate is small as $1/2\tau_\phi \sim (\delta')^2U$, which
yields a parametrically broad interval for the power-law behavior,
$(\delta')^2U \lesssim |\epsilon-\epsilon_k-\mu'|\lesssim U$. Note, 
though, that the power-law exponents $\gamma_i$ in this situation are
also small, $\gamma_i \sim \delta'$, so that the power law essentially
reduces to a logarithmic correction. 

When applied to the problem of split FES, see Eq.~(\ref{b1}), our result agrees
with that obtained by Abanin and Levitov \cite{Abanin}
for a double-step distribution. It should be 
emphasized that our approach is different from the one used in
Ref.~\onlinecite{Abanin}. While we work in the non-equilibrium 
bosonization framework and present the Green function in terms
of a single determinant $\Delta(\delta)$ at the phase $\delta=
2\pi-2\delta_0$, Abanin and Levitov  used the fermionic FES theory
and obtained the result in the form
of a product of a determinant $\Delta(-2\delta_0)$ and a Green
function, and then analyzed both terms by an approximate solution of the
corresponding Riemann-Hilbert problem. 

\begin{figure}
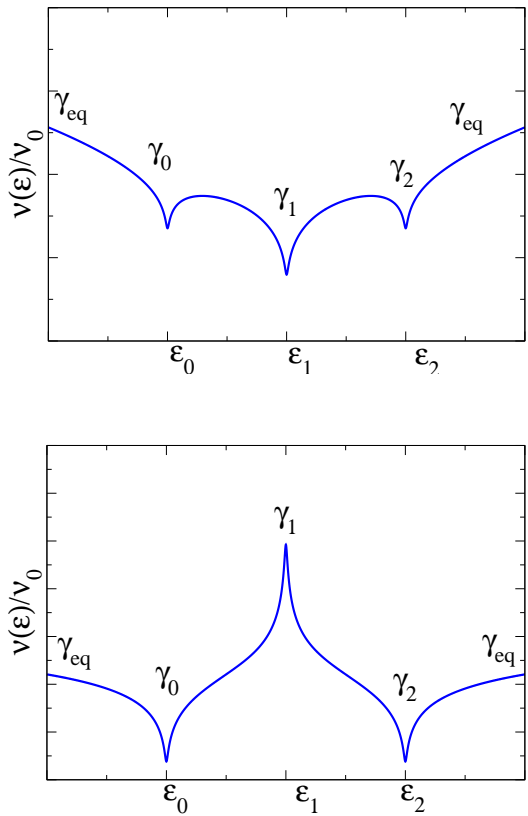

\includegraphics[width=0.8\columnwidth,angle=0]{triple_step_density_of_states_down.eps}

\vspace*{0.85cm}
\includegraphics[width=0.8\columnwidth,angle=0]{triple_step_density_of_states_down_up_down.eps}
\caption{Schematic results  for TDOS in a LL with not-too-strong
  interaction and with multiple-step
  distributions. The distributions of electrons from both reservoirs are
  assumed to be equal and of the type  shown in Fig.~\ref{figure1}A
  (upper pane) or Fig.~\ref{figure1}B (lower panel). The exponents $\gamma_i$
  characterizing ZBA at multiple edges are indicated. All
  singularities are broadened by the non-equilibrium dephasing rate
  $1/2\tau_\phi$. }
\label{figure2}
\end{figure}

Applying these results to the LL tunneling spectroscopy, see
Eq.~(\ref{d3}),  we obtain  
split power-law singularities, with modified (compared to the
equilibrium regime) exponents and with broadening
by the non-equilibrium dephasing rate $1/2\tau_\phi$. This is
illustrated in Fig.~\ref{figure2} where we show the behavior of
the tunneling density of states (TDOS), $\nu(\epsilon)=
[G^<(\epsilon)-G^>(\epsilon)]/2\pi i$.  
As discussed above, for a weak interaction the
non-equilibrium power laws reduce to logarithmic corrections and are
weakly smeared (by small $1/\tau_\phi$). For an arbitrary strength of
interaction the scale for smearing of singularities becomes comparable
to the distance between the singular points. The profile of TDOS in a
general situation can be obtained by numerical evaluation of the
Toeplitz determinant. 

It is worth mentioning some further recent works that addressed the
non-equilibrium LL spectroscopy.  Influence of
non-equilibrium conditions on exponents was also found within the
functional renormalization group approach (by using approximations justified
for weak interaction) in Ref.~\onlinecite{jakobs08}; 
dephasing was discarded there. Qualitatively similar results (modification of
exponents and oscillatory dependence of dephasing rate on the
interaction strength) were also obtained in Ref.~\onlinecite{Ngo}. The 
difference is due to the fact that the non-equilibrium setup  of
Ref.~\onlinecite{Ngo} (a biased quantum wire with an impurity inside the
interacting region)
is different from that of our work, where the non-equilibrium
distribution is assumed to be formed by scattering outside of the interacting 
LL region.

If the phase $\delta$ is close to $4\pi$, $M=2$, the leading
singularities are given by $n_k = n_l =-1$, located at $\epsilon =
\epsilon_k + \epsilon_l$, with an exponent  
$\sum_j \beta_j^2 = 2 - 2\beta_k'- 2\beta_l'+ \sum_j (\beta_j')^2$.
The next singularities are produced by $n_k=-2$, located at 
at $\epsilon = 2\epsilon_k$, with an exponent  
$\sum_j \beta_j^2 = 4 - 4\beta_k'+ \sum_j (\beta_j')^2$, etc. 
For the case of a double-step distribution and $\delta = 4\pi$ this
agrees with the exact result for $\Delta_\tau$ obtained in
Ref.~\onlinecite{GGM_long2010}.

\section{ Conclusion and Acknowledgments}

To conclude, we have considered  several  many-body  problems out of
equilibrium, including the FES, the counting statistics, and 
the tunneling spectroscopy in LL of fermions
as well as of bosons with hard-core repulsion.  
We have shown that the correlation function in all these problems  
may be  expressed  in terms of  Fredholm determinants of counting
operators. The operators are controlled by the (non-equilibrium) 
distribution function, as well as by the value of the scattering 
phase depending on the interaction strength. 
Our non-equilibrium bosonization approach
allows to solve these problems and to establish connections 
among them. 

We have performed an analysis of the long-time asymptotics  of
the relevant Fredholm determinants (which are of Toeplitz form). In the
interesting case of double-step (or, more generally, multiple-step)
distribution functions the corresponding generating functions possess
Fisher-Hartwig singularities induced by Fermi edges. 
When transformed from time into energy representation, 
the results reveal power-law behavior, 
associated with multi-particle processes at various 
discontinuities of distribution function (edges). 
The power laws thus obtained differ  from equilibrium one;
in addition, the singular behavior is broadened by non-equilibrium  
dephasing rate.

This work was supported by GIF, DFG Center for Functional Nanostructures,
Einstein Minerva Center, US-Israel
BSF, ISF, Minerva Foundation, DFG SPP 1285, EU GEOMDISS, and Rosnauka
02.740.11.5072.

\appendix

\section{Hard-core bosons at equilibrium}
\label{Tonks_girardo_equilibrium}

At equilibrium the action is quadratic and to find Green function  one
needs to calculate  
Gaussian functional integrals
 \begin{eqnarray}&&
G^>_B(\tau)=-i\rho_0\int {\cal D}\phi\exp\bigg(iS[\phi_L]+
\frac{i}{2\sqrt{2}}\sum_{\omega,q} 
\bigg[\phi_L(\omega,q)\nonumber \\&&  \times(1-e^{-i\omega
  \tau})+\bar{\phi}_L(\omega,q)(1+e^{-i\omega \tau})\bigg]+
(L\leftrightarrow R)\bigg)\,. \nonumber  
\end{eqnarray}
The action at equilibrium  is given by
\begin{eqnarray}&&
S[\phi_L]=-\sum_{\omega,q}\left(\frac{q}{2\pi}\right)^2
\bigg[\phi_L(-\omega,-q)\Pi_L^{a^{-1}}(\omega,q)\bar{\phi}_L(\omega,q)\nonumber 
\\&& 
+\frac{1}{2}\bar{\phi}_L(-\omega,-q)\Pi_L^{-1^K}(\omega,q)
\bar{\phi}_L(\omega,q)\bigg]\,. 
\end{eqnarray}
Performing the Gaussian integration over bosonic fields, one finds
the Green functions
\begin{eqnarray}&&
G^>_B(\tau)=-i\rho_0\exp\bigg(-\frac{i\pi^2}{2}
\int_{-\infty}^{\infty}(d\omega)(dq)
\bigg[\\&&
\Pi_L^K (\omega,q) 
(1-\cos\omega\tau)-2i\sin\omega\tau \Pi^r_L(\omega,q)+
(L \leftrightarrow  R)\bigg]\bigg). \nonumber 
\end{eqnarray}
The integrals over momentum $q$ can be easily calculated
\begin{eqnarray}&&
\int\frac{(dq)}{q^{2}}\Pi_L^K(\omega,q)=
\frac{i}{2\pi\omega}\coth\frac{\omega}{2T}\,, \nonumber \\&&
\int(dq)q^{-2}\Pi_L^r(\omega,q)=
-\frac{i}{4\pi\omega}\,.
\end{eqnarray}
Performing the standard integrals over $\omega$  one obtains
\begin{eqnarray}
G^>_B(\tau)=-i\rho_0\sqrt{\frac{\pi T\tau}{\sinh\pi T\tau}}
\frac{e^{-\frac{i\pi}{4}{\rm sgn} \tau}}{(1+\Lambda^2\tau^2)^{1/4}}.
\end{eqnarray}
Switching to  the frequency domain one obtains 
\begin{equation}
G^>_B(\omega)=-\frac{i\rho_0}{\pi T}\left(\frac{T}{2\Lambda}\right)^{\frac{1}{2}}
\bigg|\Gamma\left(\frac{1}{4}+\frac{i\omega}{2\pi T}\right)\bigg|^2
\exp\left(\frac{\omega}{2T}\right)\,.
\end{equation}
Similar calculation  for the $G^<$ component yield
\begin{equation}
G^<_B(\omega)=-\frac{i\rho_0}{\pi T}\left(\frac{T}{2\Lambda}\right)^{\frac{1}{2}}
\bigg|\Gamma\left(\frac{1}{4}+\frac{i\omega}{2\pi T}\right)\bigg|^2
\exp\left(-\frac{\omega}{2T}\right).
\end{equation}
The ratio between the Green functions $G^>$ and $G^<$ is equal to 
\begin{equation}
\frac{G^>_B(\omega)}{G^<_B(\omega)}=\exp{\frac{\omega}{T}}
\end{equation}
in agreement with the Fluctuation-Dissipation theorem.

\section{Mathematical background:   Szeg\H{o} and
  Fisher-Hartwig formulas} 
\label{Fisher-Hartwig-conjecture}

For completeness we present here the short summary of known
mathematical results concerning  the theory of Toeplitz determinants,
see Ref.~\onlinecite{deift09} and references therein. 
A Toeplitz matrix $\{f_{ij}\}$, $)\le i,j\le N-1$ is  generated by a
complex-valued function $f(z)$ on the unit
circle $z=e^{i\theta}$, where $\theta$ is the polar angle $\theta\in
[0,2\pi]$.  
Entries of the Toeplitz matrix are the Fourier coefficients
\begin{equation}
f_j=\int_{-\pi}^{\pi}\frac{d\theta}{2\pi} 
f\left(e^{i\theta}\right)e^{-i\theta j} \,. 
\end{equation}
The Szeg\H{o} theorem is formulated for the case when $f(z) =
e^{V(z)}$ is non-zero and sufficiently smooth on the unit circle. 
The function $V(z)$ may be described by its Fourier harmonics
\begin{equation}
V(z)=\sum_{k=-\infty}^\infty V_k z^k\,, 
\end{equation}
where
\begin{equation}
V_k=\int_{-\pi}^{\pi} \left(\frac{d\theta}{2\pi}\right) V(z) z^{-k}\,.
\end{equation}
The smoothness condition requires that $\sum_{k=-\infty}^\infty |k|
|V_k|^2$ converges. It is further assumed the $\arg V(z)$ returns to
its original value (rather than picking up a $2\pi n$ contribution with
non-zero integer $n$) when $z$ goes around the unit circle.    
According to the (strong)  Szeg\H{o} theorem, the large-$N$
asymptotic behavior of the determinant $\Delta_N[f]$ of the
corresponding matrix is  
\begin{equation}
\label{Szoego}
\Delta_N[f]=\exp\left(NV_0+\sum_{k=1}^\infty k V_kV_{-k}\right)\,.
\end{equation}

The Fisher-Hartwig formula deals with Toeplitz matrices of a more
general form, with generating function $f(z)$ having $m+1$
singularities ($m=0,1,2,\ldots$),  
\begin{equation}
\label{a7}
f(z)=e^{V(z)}z^{\sum_{j=0}^m\beta_j}\prod_{j=0}^m
\bigg|z-z_j\bigg|^{2\alpha_j}g_{z_j,\beta_j}(z)z_j^{-\beta_j}\,, 
\end{equation}
where
\begin{eqnarray}
g_{z_j,\beta_j}(z)=\left\{
\begin{array}{ll}
      e^{i\pi\beta_j} \,, & \qquad -\pi< {\rm arg} z < \theta_j  
      \\ \\
      e^{-i\pi\beta_j}  \,, & \qquad \theta_j < {\rm arg} z <\pi\,.
\end{array}
\right.
\label{jump-function}
\end{eqnarray}
The singularities are located at points
$z_j=e^{i\theta_j}$ with $j=0, \dots, m$; 
for definiteness, they can be assumed to be ordered as follows:
\begin{equation}
 -\pi=\theta_0<\theta_1< \dots< \theta_m<\pi.
\end{equation}
The strength of singularities is controlled by a set of parameters
$\alpha_j$, $\beta_j$,  satisfying
${\rm Re}\,\, \alpha_j >-\frac{1}{2}$, $\beta_j \in \mathbb{C}$.

Derivation of the asymptotic behavior of Toeplitz determinant
with Hartwig-Fisher singularities as well as overview of previous
literature can be found in the recent work \cite{deift09}. 
In this paper, we are interested in a particular case of $\alpha_j=0$
and $V(z)={\rm const}\equiv V_0$.
Indeed, the function (\ref{f_epsilon_periodic}), with 
distribution $n(\epsilon)$ having double-step or multiple-step form
(superposition of two or more zero-temperature Fermi distributions
with different chemical potentials) belongs exactly to this class of
function. We thus present the results for this particular case only,
referring the reader to Ref.~\onlinecite{deift09} for general results. 
If all $\beta_j$ are sufficiently close to each other, such that
$|{\rm Re} \beta_j- {\rm Re} \beta_k| < 1$ for all $j,k=0,\ldots,m$,
the asymptotic behavior of the determinant reads
\begin{eqnarray}   
\label{fisher-hartwig}
\Delta_N &=& e^{NV_0} N^{-\sum_{j=0}^m \beta_j^2} \prod_{0\le j<k\le m}
|z_j-z_k|^{2\beta_i\beta_k} \nonumber \\ &\times&
 \prod_{j=0}^m G(1+\beta_j)G(1-\beta_j)\ ,
\end{eqnarray}
where $G(z)$ is the Barnes $G$-function. 
A more general results that is valid for any values of $\beta_j$ and
yields also subleading contributions has the following form:
\begin{eqnarray}   
\label{fisher-hartwig-general}
\Delta_N &=& e^{NV_0} \sum_{n_0+\ldots+n_m=0}\ \prod_{j=0}^m z_j^{n_jN}
\nonumber \\
&\times &  \left[ N^{-\sum_{j=0}^m \beta_j^2} \prod_{0\le j<k\le m}
|z_j-z_k|^{2\beta_i\beta_k} \right. \nonumber \\
&\times & \left. 
 \prod_{j=0}^m G(1+\beta_j)G(1-\beta_j)\right]_{\beta_j \to
\beta_j + n_j}\,.
\end{eqnarray}
The summation in eq.~(\ref{fisher-hartwig-general}) goes over all sets
of integers $n_0, n_1,\ldots,n_m$ satisfying $\sum_{j=0}^m n_j
=0$. This formula plays a central role in our analysis of the
asymptotics of determinants governing the Green functions of many-body
problems in Sec.~\ref{s:Fisher-Hartwig}. As we show there, the shifts
by integers $n_j$ in Eq.~(\ref{fisher-hartwig-general}) generate
contributions corresponding to multiple Fermi edges to these Green
functions.


\begin{thebibliography}{99}

\bibitem{Anderson}  P.W. Anderson, Phys. Rev. Lett. {\bf 18}, 1049 (1967).


\bibitem{Nozieres} P.\,Nozi\`eres and C.\,T.\,De Dominicis,
{\it Phys. Rev.} {\bf 178}, 1097 (1969).

\bibitem{Tomonago} S. Tomonaga, Prog. Theor. Phys. {\bf 5}, 544 (1950).

\bibitem{Luther} A. Luther and I. Peschel, Phys. Rev. B {\bf 9}, 2911 (1974).


\bibitem{Kondo} J. Kondo, Prog. Theor. Phys. {\bf 32}, 37  (1964).

\bibitem{Yuval} G. Yuval and P. W. Anderson,  Phys. Rev. B {\bf 1}, 1522 (1970).


\bibitem{GGM_long2010} D.B. Gutman, Y. Gefen, and A.D. Mirlin,
  Europhys. Letters {\bf 90}, 37003 (2010); Phys. Rev. B {\bf 81},
  085436 (2010).   

\bibitem{GGM_short2010} D.B. Gutman, Y. Gefen, and A.D. Mirlin,
  arxiv:1003.5433.  





\bibitem{stone} M. Stone, \textit{Bosonization} (World Scientific, 1994).

\bibitem{Delft} J. von Delft and  H. Schoeller,
Annalen Phys. {\bf 7}, 225  (1998).

\bibitem{Gogolin}
A.O. Gogolin, A.A. Nersesyan, and A.M. Tsvelik,
\textit{Bo\-son\-iza\-tion in Strongly Correlated Systems},
(University Press, Cambridge 1998).

\bibitem{giamarchi}
T. Giamarchi, \textit{Quantum Physics in One Dimension},  (Claverdon
Press Oxford, 2004).

\bibitem{maslov-lectures} D.L.~Maslov, in {\it Nanophysics: Coherence
    and Transport}, edited by H.~Bouchiat, Y.~Gefen, G.~Montambaux,
  and J.~Dalibard (Elsevier, 2005), p.1.


\bibitem{Kamenev} for review of the Keldysh technique
see, e.g., J.~Rammer and H.~Smith,
Rev. Mod. Phys. {\bf 58}, 323 (1986);
A. Kamenev, in {\it
Nanophysics: Coherence and Transport} (Elsevier,  2005), edited by
H.~Bouchiat, Y.~Gefen, G.~Montambaux,  and J.~Dalibard, p. 177;
A.~Kamenev and A.~Levchenko, Adv. Phys. {\bf 58}, 197 (2009). 


\bibitem{Schotte} K.D.~Schotte and U.~Schotte, Phys. Rev. B {\bf 182},
  479 (1969). 

\bibitem{Abanin}
D.A. Abanin and L.S. Levitov,  Phys. Rev. Lett. {\bf  93},  126802 (2004);
D.A. Abanin and L.S. Levitov,  Phys. Rev. Lett. {\bf 94},   186803 (2005).

\bibitem{tunnel-spectroscopy}
Y.-F.~Chen, T.~Dirks, G.~Al-Zoubi, N.~Birge, and N.~Mason
Phys. Rev. Lett. {\bf 102}, 036804 (2009). 

\bibitem{tunnel-spectroscopy-qhe} C. Altimiras, H. le Sueur,
U. Gennser, A. Cavanna, D. Mailly, F. Pierre,
Nature Physics {\bf 6}, 34 (2010). 

\bibitem{jakobs08} S.G.~Jakobs, V.~Meden, and H.~Schoeller,
Phys. Rev. Lett. {\bf 99}, 150603 (2007). 

\bibitem{gutman08} D. B. Gutman, Y.~Gefen, and A. D. Mirlin
Phys. Rev. Lett. {\bf 101}, 126802 (2008); Phys. Rev. B {\bf 80},
045106 (2009).  

\bibitem{trushin08} 
M. Trushin and A. L. Chudnovskiy, Europhys. Letters,  {\bf 82}, 17008
(2008).

\bibitem{pugnetti09}
S. Pugnetti, F. Dolcini, D. Bercioux, and H. Grabert, Phys. Rev. B 79,
035121 (2009). 

\bibitem{Ngo} 
S.~Ngo Dinh, D.A.~Bagrets, and A.D.~Mirlin, Phys. Rev. B {\bf 81},
081306 (R) (2010). 

\bibitem{takei10} S. Takei, M. Milletarì, and B. Rosenow, 
Phys. Rev. B {\bf 82}, 041306(R) (2010).

\bibitem{bena10} C. Bena, Phys. Rev. B {\bf 82}, 035312 (2010). 

\bibitem{Maslov} D.L. Maslov and M. Stone, Phys. Rev. B {\bf 52},
  R5539 (1995).

\bibitem{Safi}I.~Safi and H.J.~Schulz, Phys. Rev. B {\bf 52}, R17040 (1995).

\bibitem{Ponomarenko}  V.V.~Ponomarenko, Phys. Rev. B {\bf 52},
  R8666 (1995).

\bibitem{Safi97} I.~Safi, Ann. Phys. {\bf 22}, 463 (1997).

\bibitem{lehur}
K. Le Hur, Phys. Rev. B {\bf 65}, 233314 (2002);
Phys. Rev. Lett. {\bf 95}, 076801 (2005);
  Phys. Rev. B {\bf 74}, 165104 (2006).

\bibitem{Pham00} K.-V.~Pham, 
  M.~Gabay, and P.~Lederer, Phys. Rev. B {\bf 61}, 16397 (2000)

\bibitem{LeHur08} K. Le Hur, B. I. Halperin, and
  A. Yacoby, Ann. Phys. {\bf 323}, 3037 (2008);
H.~Steinberg, G.~Barak, A.~Yacoby, L.N.~Pfeiffer, K.W.~West,
B.I.~Halperin and K.~Le Hur, Nat. Phys. {\bf 4}, 116 (2008).

\bibitem{Berg09}
E. Berg, Y. Oreg, E.-A. Kim, and F. von Oppen,  
Phys. Rev. Lett. {\bf 102}, 236402 (2009).

\bibitem{Deshpande10}
V.V.~Deshpande, M.~Bockrath, L.I.~Glazman, and A.~Yacoby,
Nature {\bf 464}, 209 (2010). 

\bibitem{Chalker07} J.T.~Chalker, Y.~Gefen, and M.Y.~Veillette,
Phys. Rev. B {\bf 76}, 085320 (2007).

\bibitem{Levkivskyi09} I.P.~Levkivskyi and E.V.~Sukhorukov,
Phys. Rev. Lett. {\bf 103} 036801 (2009).

\bibitem{Kovrizhin10} D. L. Kovrizhin and J. T. Chalker,
Phys. Rev. B {\bf 81}, 155318 (2010); arXiv:1009.4555.

\bibitem{Kim07} N.Y.~Kim, P.~Recher, W.D.~Oliver, Y.~Yamamoto,
  J.~Kong,  and H.~Dai,  Phys. Rev. Lett. {\bf 99}, 036802 (2007).

\bibitem{Wu07}
F.~Wu, P.~Queipo, A.~Nasibulin, T.~Tsuneta, T.H.~Wang, E.~Kauppinen,
and P.J.~Hakonen,   Phys. Rev. Lett. {\bf 99}, 156803 (2007).

\bibitem{Levitov-noise} 
L.S. Levitov and  G.B. Lesovik, JETP Lett. {\bf 58}, 230 (1993);
L.S. Levitov, H. Lee, and G.B. Lesovik,
J. of Math. Phys. {\bf 37}, 4845 (1996);
D.A. Ivanov, H.~Lee, and L.S.~Levitov,
Phys. Rev. B {\bf 56}, 6839 (1997). 

\bibitem{cold_atoms} 
S. Hofferberth, I. Lesanovsky, T. Schumm, A. Imambekov, V. Gritsev,
E. Demler, and  J. Schmiedmayer, Nature Physics {\bf 4}, 489 (2008).

\bibitem{Girardeau}
L. Tonks, Phys. Rev. {\bf 50}, 955 (1936); 
M.?D. Girardeau, J. Math. Phys. (N.Y.) {\bf 1}, 516 (1960).

\bibitem{Haldane} F.D.M. Haldane,  J. Phys.C {\bf 14}, 2585 (1981). 

\bibitem{Klich}
I. Klich, 
in \emph{Quantum Noise in Mesoscopic Systems},
ed. by Yu. V. Nazarov (Kluwer, Dordrecht, 2003);
cond-mat/0209642.
 
\bibitem{Muzykantskii03} 
B.A. Muzykantskii and Y. Adamov,
Phys. Rev. B {\bf 68}, 155304 (2003).

\bibitem{Shelankov03}  
A. Shelankov and J. Rammer,
Europhysics Letters {\bf 63}, 485 (2003). 

\bibitem{Braunecker06} 
B.~Braunecker, Phys. Rev. B 73, 075122 (2006). 

\bibitem{Schoenhammer07} 
K. Schoenhammer, Phys. Rev. B {\bf 75}, 205329 (2007).

\bibitem{Avron08} 
J.E.~Avron, S.~Bachmann, G.M.~Graf, and I.~Klich,
Commun. Math. Phys. {\bf 280}, 807 (2008).

\bibitem{Hassler08} 
F.~Hassler, M.V.~Suslov, G.M.~Graf, M.V.~Lebedev, G.B.~Lesovik, and G. Blatter,
Phys. Rev. B {\bf 78}, 165330 (2008).  

\bibitem{Abanov09}
A.G. Abanov and D. A. Ivanov, Phys. Rev. B 79, 205315 (2009).  

\bibitem{Jimbo80}
M.~Jimbo, T.~Miwa, Y.~M\^ori,  and M.~Sato,
Physica D {\bf 1}, 80 (1980).

\bibitem{Izergin98}
A.G.~Izergin and A.G.~Pronko,
Nuclear Physics B {\bf 520 [FS]}, 594 (1998).

\bibitem{bettelheim06}
E.~Bettelheim, A.G.~Abanov, P.~Wiegmann,
Phys. Rev. Lett. {\bf 97}, 246402 (2006).

\bibitem{Zvonarev}
M. B. Zvonarev, V. V. Cheianov, and T. Giamarchi,
J. Stat. Mech. P07035 (2009).

\bibitem{Nazarov}
I. Snyman and Y.V. Nazarov, Phys. Rev. Lett. {\bf 99}, 096802 (2007).

\bibitem{Marquardt}
I. Neder and F. Marquardt, New Journal of Physics {\bf 9}, 112 (2007).


\bibitem{deift09} P.~Deift, A.~Its, and I.~Krasovsky,
  arXiv:0905.0443. 


\end{thebibliography}
\end{document}